\documentclass[aps,prl,twocolumn,superscriptaddress]{revtex4-2}
\usepackage{bm}                 % bold math
\usepackage{graphicx}
\usepackage{xcolor}
\usepackage{float}
\usepackage{amsmath}
\newcommand{\Chula}{Department of Physics, Faculty of Science, Chulalongkorn University, Patumwan, Bangkok 10330, Thailand}

\begin{document}
%%%%%%%%%%%%%%%%%%%%%%%%%%%%%%%%%%%%%%%%%%%%%%%%%%%%%%%%%%%%%%%%%%%%%%%%
%   document header                                                    %
%%%%%%%%%%%%%%%%%%%%%%%%%%%%%%%%%%%%%%%%%%%%%%%%%%%%%%%%%%%%%%%%%%%%%%%%
%\title[Edge conduction in 4LG electron-hole system]{Edge conduction in Bernal-stacked tetralayer graphene electron-hole systems}
\title{Probing the Anisotropic Fermi Surface in Tetralayer Graphene via Transverse Magnetic Focusing}

\author{Illias Klanurak}
\affiliation{\Chula}

\author{Kenji Watanabe}
\affiliation{Research Center for Electronic and Optical Materials, National Institute for Materials Science, 1-1 Namiki, Tsukuba 305-0044, Japan}

\author{Takashi Taniguchi}
\affiliation{Research Center for Materials Nanoarchitectonics, National Institute for Materials Science,  1-1 Namiki, Tsukuba 305-0044, Japan}

\author{Sojiphong Chatraphorn}
\affiliation{\Chula}

\author{Thiti Taychatanapat}
\email[]{thiti.t@chula.ac.th}
\affiliation{\Chula}

\date{\today}
\begin{abstract}
	  Bernal-stacked tetralayer graphene (4LG) exhibits intriguing low-energy properties, featuring two massive subbands and showcasing diverse features of topologically distinct, anisotropic Fermi surfaces, including Lifshitz transitions and trigonal warping.  Here, we study the influence of the band structure on electron dynamics within 4LG using transverse magnetic focusing. Our analysis reveals two distinct focusing peaks corresponding to the two subbands. Furthermore, we uncover a pronounced dependence of the focusing spectra on crystal orientations, indicative of an anisotropic Fermi surface. Utilizing the semiclassical model, we attribute this orientation-dependent behavior to the trigonal warping of the band structure. This phenomenon leads to variations in electron trajectories based on crystal orientation. Our findings not only enhance our understanding of the dynamics of electrons in 4LG, but also offer a promising method for probing anisotropic Fermi surfaces in other materials. 
\end{abstract}

\maketitle

The distinct electronic properties of multilayer graphene, in contrast to monolayer graphene, are shaped by unique features \cite{mccann2013electronic,craciun2009trilayer,datta2017strong,hirahara2018multilayer}. Trigonal warping (TW) plays a pivotal role in this differentiation, inducing a distortion of the circular Fermi surface near $K$ and $K'$ points in the graphene Brillouin zone due to interlayer interactions \cite{charlier1991first}. This effect introduces a discrete 3-fold rotational symmetry, breaking the continuous rotational symmetry of the Fermi surface and causing anisotropy in the low-energy band structure \cite{charlier1991first}. Numerous studies have emphasized the significance of TW, associating it with diverse phenomena such as correlated phases in bilayer graphene \cite{seiler2022quantum}, emergent Dirac gullies \cite{zibrov2018emergent}, Landau level hybridization in trilayer graphene \cite{datta2018landau}, Lifshitz transitions in Bernal-stacked tetralayer graphene (4LG) \cite{shi2018tunable}, and topological phase transitions in twisted double bilayer graphene \cite{mohan2021trigonal}. Recognizing the importance of the TW effect is crucial for anticipating the transport characteristics of materials

Although scanning tunneling microscopy has facilitated quasi-direct observation of the TW effect in graphene \cite{joucken2020determination}, its elucidation in transport experiments remains challenging. Prior methods, including ballistic transport in bilayer graphene with superimposed antidot arrays \cite{oka2019ballistic} and quantum point contacts \cite{ingla2023specular}, have been employed. However, these approaches could introduce inhomogeneity that can mask the intrinsic transport characteristics of bilayer graphene~\cite{Lee2022}. In our study, we leverage transverse magnetic focusing (TMF) on 4LG to investigate the low-energy anisotropic band structure associated with TW. By applying a transverse magnetic field to guide electron trajectories, TMF enables us to derive insights into the electronic properties of materials~\cite{van1989coherent,rokhinson2004spin,taychatanapat2013electrically, Chen2016, lee2016ballistic, Lo2017, berdyugin2020minibands, Gupta2021, Rao2023}. 

\begin{figure*}
  \includegraphics{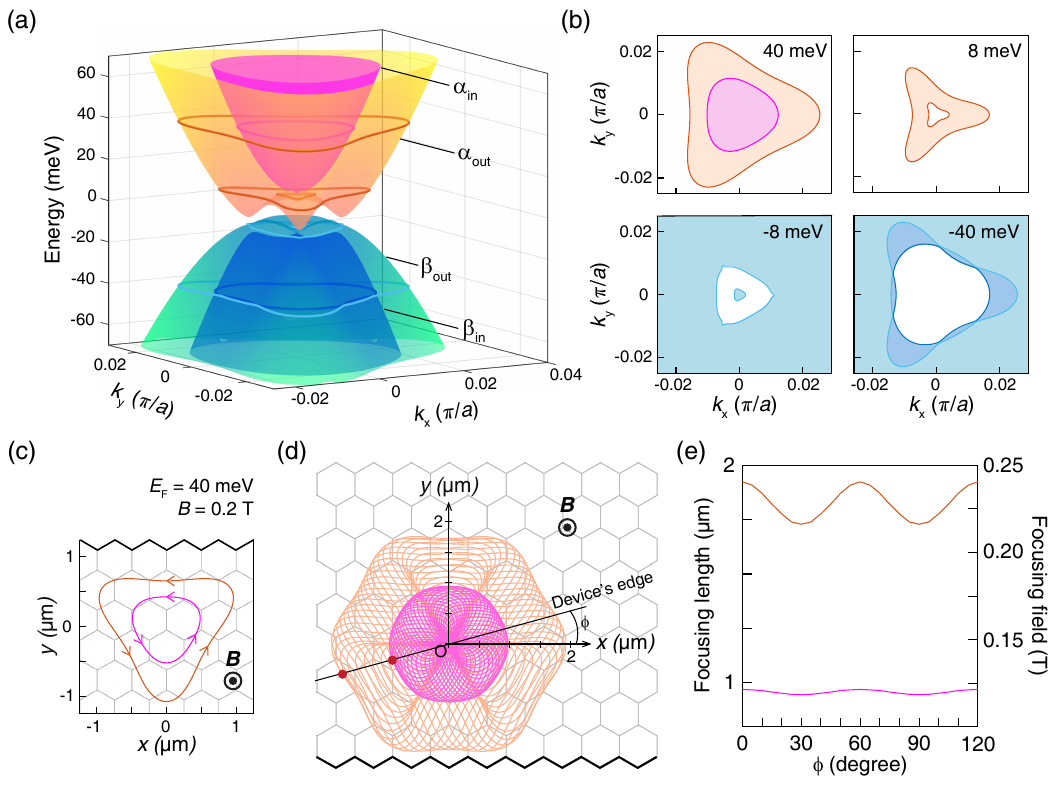}
  \caption{(a) Electronic band structure of 4LG. The inner and outer conduction bands are labeled by $\alpha_{\mathrm{in}}$ and $\alpha_{\mathrm{out}}$ while the inner and outer valence bands are labeled by $\beta_{\mathrm{in}}$ and $\beta_{\mathrm{out}}$. (b) Fermi surfaces of 4LG at 4 different Fermi energies. The shading indicates filled states for each band. (c) Electron trajectories in real space at $E_{\mathrm{F}}$ = 40 meV and magnetic field of $0.2$~T. The pink and orange trajectories corresponds to electrons from $\alpha_{\mathrm{in}}$ and $\alpha_{\mathrm{out}}$ bands, respectively. (d) Simulated trajectories of electrons injected through point O in all directions under the same conditions as in (c). The angle $\phi$ measures the orientation of the device's edge relative to the zigzag edge of 4LG. The red dots indicate the focusing points at which the caustic lines intersect the device's edge. (e) Focusing length at $0.2$~T and focusing field for focusing length of $1.6$~$\mu$m as a function of $\phi$ for electron trajectories in (d).}
\end{figure*}

The advantage of using 4LG in this study lies in its band structure, which hosts charge carriers with two distinct effective masses \cite{koshino2011landau,klanurak2022magnetoconductance}. We denote the inner and outer electron (hole) bands as $\alpha_{\mathrm{in}}$ ($\beta_{\mathrm{in}}$) and $\alpha_{\mathrm{out}}$ ($\beta_{\mathrm{out}}$), respectively, as illustrated in Figure 1a. The band structure of 4LG is parameterized by the Slonczewski-Weiss-McClure (SWMcC) parameters of graphite \cite{charlier1991first} ($\gamma_{0}$-$\gamma_{5}$ and $\delta$). Notably, the interlayer hopping parameter between non-dimer sites, $\gamma_{3}$, is identified as a key contributor to the TW effect in multilayer graphene \cite{charlier1991first}. 

The influence of $\gamma_{3}$ on each band at specific Fermi energies $E_{\mathrm{F}}$ can be visualized by the shape of Fermi surfaces, as shown in Figure 1b. For energy $|\epsilon(\mathbf{k})|>8$ meV, the Fermi surfaces at each $K$ point consist of two pockets, with those from $\alpha_{\mathrm{out}}$ and $\beta_{\mathrm{out}}$ bands exhibiting more distortion than those of $\alpha_{\mathrm{in}}$ and $\beta_{\mathrm{in}}$ bands. We propose that the distinct degrees of TW between the inner and outer bands can be discerned through ballistic magnetotransport measurements utilizing the TMF technique. 

In TMF technique, we inject charge carriers into a two-dimensional electron gas subjected to a small perpendicular magnetic field $B$. Employing the semiclassical framework\cite{ashcroft2022solid}, the trajectories of charge carriers are dictated by the energy dispersion $\epsilon(\mathbf{k})$ at $E_{\mathrm{F}}$. The evolution of position $\mathbf{r}$ and momentum $\hbar \mathbf{k}$ for the carriers is governed by the semiclassical equation:

\[\dot{\mathbf{r}} = \frac{1}{\hbar} \frac{\mathrm{d}\epsilon}{\mathrm{d}\mathbf{k}} \ , \  \hbar \dot{\mathbf{k}} = -e\dot{\mathbf{r}} \times B \hat{z} \] 

where $\hbar$ is the reduced Planck constant and $e$ is the elementary charge.  In essence, the real-space trajectory of charge carriers mirrors the shape of their Fermi surfaces but is scaled by $\hbar/eB$ and rotated by 90$^{\circ}$. As depicted in Figure 1c, electron trajectories in real space at $E_{\mathrm{F}}=40$ meV under $B=0.2$ T are illustrated, with the $x$-axis aligned with the zigzag edge of 4LG. Electrons within each pocket trace closed orbits with their direction determined by the Lorentz force. Notably, electrons from the $\alpha_{\mathrm{in}}$ pocket (pink-color path) exhibit smaller orbits in real space compared to those in the $\alpha_{\mathrm{out}}$ pocket (orange-color path) due to the smaller size of their Fermi surface.

In Figure 1d, we depict the trajectories of electrons emanating from a point $O$ in various directions, all under identical energy and magnetic field conditions as those shown in Figure 1c. This illustration captures how each set of trajectories within a band collectively evolves into a caustic, a zone characterized by a significantly high concentration of carriers. This results in a pronounced focusing pathway for these carriers. Notably, the $\alpha_{\mathrm{out}}$ caustic exhibits heightened sensitivity to TW, leading to its adoption of a hexagonal shape, in contrast to the $\alpha_{\mathrm{in}}$ caustic, which retains a more circular form. The caustic lines for each band converge on different focusing points along the device's edge, marked as red dots in Figure 1d. 

Additionally, as a result of an anisotropic Fermi surface, we observe variations in the focusing length as a function of device orientation, denoted by $\phi$, relative to the zigzag edge. This dependency is clearly demonstrated in Figure 1e for both the $\alpha_{\mathrm{in}}$ and $\alpha_{\mathrm{out}}$ bands. Specifically, Figure 1e reveals that the variability in focusing length with respect to $\phi$ is more pronounced for the $\alpha_{\mathrm{out}}$ band, whereas it remains relatively stable for the $\alpha_{\mathrm{in}}$ band.  Notably, device orientations aligned with the zigzag edge ($\phi=0^{\circ}$) and the armchair edge ($\phi=90^{\circ}$) yield the maximum difference in focusing length and, consequently, focusing field, as illustrated in Figure 1e. 

\begin{figure*}
  \includegraphics{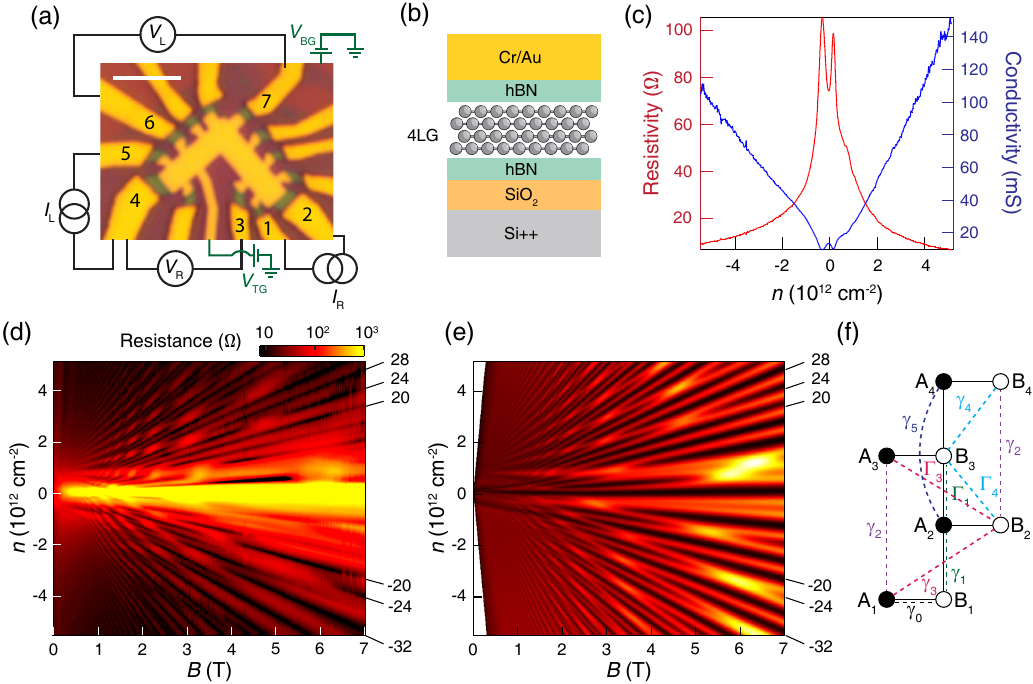}
  \caption{ (a) Optical image of the 4LG device with nonlocal TMF measurement configurations, where each contact is labeled by numbers 1-7. The scale bar is 2 $\mu$m. (b) Schematic diagram of device cross-section. We encapsulate a 4LG flake with top and bottom hBN flakes. Highly doped silicon and Cr/Au are used as bottom and top gates. (c) Resistivity (red line) and conductivity (blue line) as functions of $n$ at $2.4$~K from the left side of the device. (d) Longitudinal resistance as a function of $B$ and $n$. (e) Calculated density of states of 4LG. The labeled numbers indicate filling factors of gap states (black lines). (f) Crystal structure of 4LG and its hopping parameters.}
\end{figure*}

Therefore, to investigate the anisotropy of the Fermi surfaces in 4LG with TMF, we fabricate an L-shaped device, illustrated in Figure 2a. To ensure the maximum variation of the TMF signal, the left and right sides of the device are etched in parallel to the cleaved lines of the exfoliated 4LG flake, which is expected to exhibit either zigzag or armchair edges (see Supporting Information for details) \cite{neubeck2010direct,almeida2011identification}.

To explore the ballistic behavior of electrons in 4LG, we construct hBN/4LG/hBN heterostructure using the dry transfer technique \cite{wang2013one}, depicted in Figure 2b. The heterostructure is then etched in $\mathrm{O_{2}/CHF_{3}}$ plasma, while electron beam lithography along with thermal evaporation of Cr/Au are employed to define top gate and electrical contacts. Both top and back gates are utilized to independently control carrier density $n$ and electric displacement field $D$.  Each contact is referred to by the number labels shown in Figure 2a. All the measurements in this study are conducted at 2.4 K using the standard lock-in technique with an excitation frequency of 17 Hz. We maintain $D$ at zero to preserve the intrinsic band structure of 4LG.

We initially evaluate the device quality by examining the longitudinal resistivity as a function of $n$, as illustrated in Figure 2c (see Supporting Information for details). The $n$ value is determined from the period of Shubnikov-de Haas oscillations at a finite magnetic field. The resistivity profile reveals two distinct peaks around the charge neutrality point (CNP), along with a small shoulder at positive $n$, indicative of 4LG~\cite{shi2018tunable}. These features coincide precisely with Lifshitz transition points, which correspond to the local maxima or minima in the band structure. At large $n$, the conductivity demonstrates linear behavior, corresponding to field-effect mobilities of approximately 180,000 and 110,000 $\mathrm{cm^{2}V^{-1}s^{-1}}$ for electrons and holes, respectively. These values are comparable to prior studies on ballistic transport in graphene \cite{mayorov2011micrometer,taychatanapat2013electrically,sandner2015ballistic,oka2019ballistic}.

To obtain an accurate description of the band structure, we utilize Landau level (LL) crossings to determine the values of the  SWMcC parameters \cite{Taychatanapat2011}. In Figure 2d, the longitudinal resistance between contacts 5 and 6 is presented as a function of $B$ and $n$. The LL crossings manifest as bright color spots in Figure 2d. By comparing the LL crossings in the simulated density of states (Figure 2e) with experimental data, we achieve excellent agreement using the SWMcC parameters reported in Ref. \citenum{shi2018tunable} ($\gamma_{0}=3$, $\gamma_{1}=0.39$, $\gamma_{2}=-0.02$, $\gamma_{3}=0.3$, $\gamma_{4}=0.04$, $\gamma_{5}=0.04$, $\Gamma_{1}=0.32$, $\Gamma_{3}=0.25$, $\Gamma_{4}=0.032$, and $\delta=0.041$ eV). Additional parameters $\Gamma_{1}$, $\Gamma_{3}$, and $\Gamma_{4}$ are introduced to accommodate interlayer hopping between bulk layers (see Figure 2f), reflecting differences in interlayer interactions between the bulk and surfaces of 4LG \cite{wu2015detection}. 

\begin{figure*}
  \includegraphics{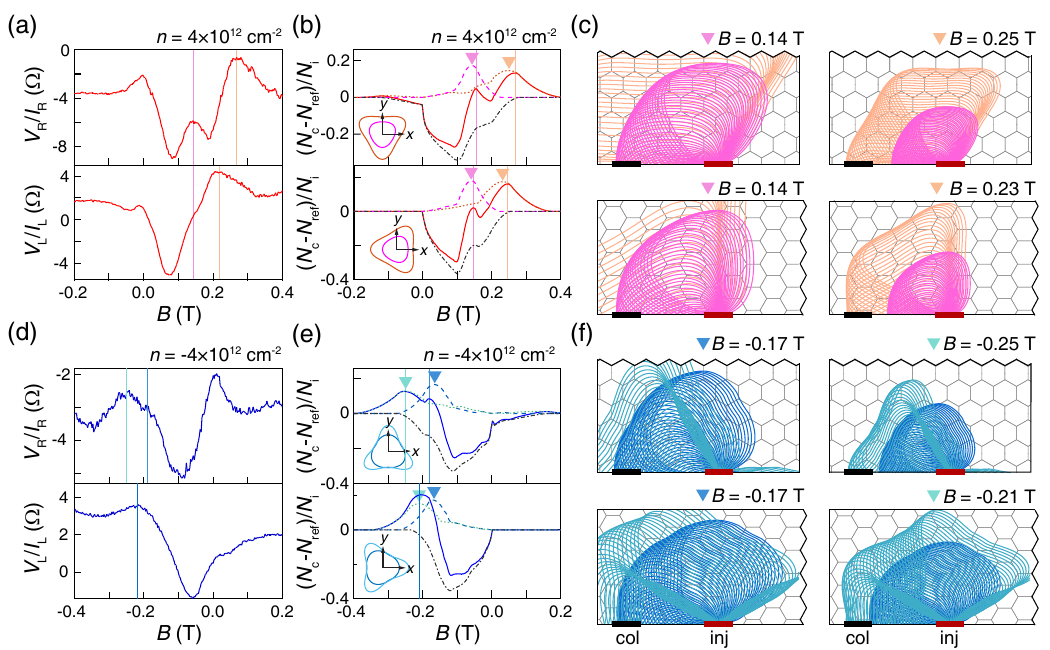}
  \caption{(a--c) Comparison of measure and simulated TMF spectra at $n = 4\times10^{12}$~$\mathrm{cm^{-2}}$: (a) The measured TMF spectra at $2.4$~K with vertical lines indicating the focusing peaks. (b) The simulated TMF spectra (solid line) alongside ratios of $N_{\mathrm{c}}/N_{\mathrm{i}}$ for $\alpha_{\mathrm{in}}$ and $\alpha_{\mathrm{out}}$ bands (pink dashed lines and orange dotted lines, respectively). The black dashed-dotted lines represent the combined $-N_{\mathrm{ref}}/N_{\mathrm{i}}$ from both bands. (c) The simulated trajectories in real space at magnetic fields corresponding to the peak positions in (b). The red and black thick horizontal lines mark the positions of injectors and collectors, respectively. The top and bottom rows of these panels correspond to data from the right and left sides of the device, respectively. (d--e) Comparison of measured and simulated TMF spectra at $n = -4\times10^{12}$~$\mathrm{cm^{-2}}$: (d) The measured TMF spectra at $2.4$~K. (e) The simulated TMF spectra (solid line) and the ratios of $N_{\mathrm{c}}/N_{\mathrm{i}}$  for $\beta_{\mathrm{in}}$ and $\beta_{\mathrm{out}}$ bands (blue dashed lines and teal dotted dashed lines, respectively). (f) The simulated trajectories in real space.}
\end{figure*}

We then perform nonlocal measurement to obtain TMF spectra for both sides of the device. Figure 2a illustrates the specific measurement configurations employed. Electron injection and collection are carried out through contacts 5 (1) and 6 (3) for the left (right) side, respectively, with a distance of $1.6$~$\mathrm{\mu m}$ between the injector and collector on both sides. The resulting TMF spectra for electrons at $n = 4\times10^{12}$ $\mathrm{cm^{-2}}$, presented in Figure 3a as a function of $B$, reveal two distinct peaks at positive $B$, as indicated by vertical lines. These peaks, denoted as $B_{\mathrm{f}}$, correspond to the magnetic fields required to focus electrons onto the collector. Notably, lower $B_{\mathrm{f}}$ values are associated with the $\alpha_{\mathrm{in}}$ band, while higher values are linked to the $\alpha_{\mathrm{out}}$ band. This difference in peak values is attributed to the larger Fermi surface of the $\alpha_{\mathrm{out}}$ band compared to the $\alpha_{\mathrm{in}}$ band, necessitating a stronger magnetic field for electron focusing at the same distance. 

Significantly, distinctions emerge in the TMF spectra between the two sides. The $B_{\mathrm{f}}$ values for $\alpha_{\mathrm{in}}$ electrons remain consistent on both sides, aligning with their nearly circular caustic curve (see Figure 1d). Conversely, for the $\alpha_{\mathrm{out}}$ band, variations in $B_{\mathrm{f}}$ values become apparent between the two sides, despite the equal distances between the injector and collector on both sides.  The high mobility inherent in our device suggests that these discrepancies in $B_{\mathrm{f}}$ are not a consequence of disorder. Further, quantum Hall measurements for both sides yield identical results, confirming uniform Bernal stacking across the device (see Supporting Information for details).  These observations collectively point to the anisotropic Fermi surface as the primary contributor to the observed disparity, a consequence of TW.

To confirm our assumption, we perform numerical calculations to determine TMF spectra by tracking the count of carriers reaching the collector~\cite{lee2016ballistic,berdyugin2020minibands}. The trajectories of carriers are extracted utilizing the semiclassical model from the calculated Fermi surfaces at the specified $n$. These Fermi surfaces are obtained from the band structure of 4LG, incorporating SWMcC parameters derived from quantum Hall data (see Supporting Information for details). The nonlocal resistance, represented as $V_{\mathrm{R}}/I_{\mathrm{R}}$, is approximated through the ratio $(N_{\mathrm{c}}-N_{\mathrm{ref}})/N_{\mathrm{i}}$,  where $N_{\mathrm{i}}$ is the number of injected carriers, $N_{\mathrm{c}}$ is the number of collected carriers, and $N_{\mathrm{ref}}$ is the number of carriers accumulated at the other reference voltage probe. We weigh the number of injected carriers in each pocket by their electronic density of states (see Supporting Information for details). In our simulation, we neglect the specular reflection of electrons at the device's edge, given the absence of higher-order peaks in our TMF data. The resulting simulation, depicted by the solid lines in Figure 3b, reveals that the best-fit values of $\phi$ for the right-side (RS) and left-side (LS) contacts are $0^{\circ}$ (zigzag edge) and $90^{\circ}$ (armchair edge), respectively. These values align with the device's design, expected to correspond to the zigzag or armchair edges of 4LG. The pink dashed lines and orange dotted lines in Figure 3b illustrate the ratio $N_{\mathrm{c}}/N_{\mathrm{i}}$ for electrons from the $\alpha_{\mathrm{in}}$ and $\alpha_{\mathrm{out}}$ bands, respectively. Additionally, the black dashed-dotted line represents the combined ratio $-N_{\mathrm{ref}}/N_{\mathrm{i}}$ from both bands which provides a negative background in our TMF spectra.

To elucidate the physical significance of each peak, we compute the trajectories of the ensemble of injected electrons at each $B_{\mathrm{f}}$, represented by pink (for $\alpha_{\mathrm{in}}$) and orange (for $\alpha_{\mathrm{out}}$) triangles in Figure 3b. The outcomes are presented in Figure 3c, where zigzag lines denote the orientation of the zigzag edge of 4LG. The simulation unveils that $B_{\mathrm{f}}$ for each band aligns with the magnetic field strength that focuses the maximum number of electrons onto the collector, as depicted by the caustic in Figure 3c. Moreover, it highlights that the peak positions for $\alpha_{\mathrm{in}}$ electrons remain consistent for both sides, while the $\alpha_{\mathrm{out}}$ peak for the LS contacts shifts towards lower $B$ compared to the RS contacts. This behavior is attributed to the more pronounced TW effect experienced by the $\alpha_{\mathrm{out}}$ electrons in contrast to the $\alpha_{\mathrm{in}}$ electrons, evident from the Fermi surfaces.

While most peak positions in the simulation generally exhibit excellent alignment with the experimental data, a noticeable discrepancy arises in the actual TMF data for the left side of the device. Here, the peak corresponding to the $\alpha_{\mathrm{in}}$ band manifests as a subtle shoulder rather than a distinct peak, in contrast to the simulation results (Figure 3a and 3b). In addition, the focusing field for the $\alpha_{\mathrm{out}}$ band appears smaller than in simulation. This disparity is likely rooted in the inherent simplicity of the model employed for calculating trajectories and background signal. To enhance the precision of our analysis, more advanced methodologies, such as the Landauer-Büttiker scattering theory \cite{beconcini2016scaling, Petrovi2017}, could be employed to offer a more accurate portrayal of the experimental observations. Moreover, considering the charge accumulation at boundaries resulting from graphene's finite size, which modifies carrier trajectories and impacts the focusing field, can enhance accuracy of the simulation~\cite{taychatanapat2013electrically,Silvestrov2008}.   However, pursuing such refinements extends beyond the scope of the present paper.

%While the peak positions in the simulation generally exhibit excellent alignment with the experimental data, a noticeable discrepancy arises in the actual TMF data for the left side of the device. Here, the peak corresponding to the $\alpha_{\mathrm{in}}$ band manifests as a subtle shoulder rather than a distinct peak, in contrast to the simulation results (Figure 3a and 3b). This disparity is likely rooted in the inherent simplicity of the model employed for calculating trajectories and background signal. To enhance the precision of our analysis, more advanced methodologies, such as the Landauer-Büttiker scattering theory \cite{beconcini2016scaling, Petrovi2017}, could be employed to offer a more accurate portrayal of the experimental observations. However, pursuing such refinements extends beyond the scope of the present paper.

Next, we analyze the TMF spectra for holes at $n = -4\times10^{12}$~cm$^{-2}$  utilizing the identical measurement configurations employed for electrons. The resulting spectra are illustrated in Figure 3d. Notably, the focusing peaks for holes manifest at negative values of $B$, aligning with expectations as holes possess opposite charges to electrons. Consequently, they necessitate the opposite sign of $B$ to guide them toward the collectors. Similar to the electron regime, the spectra for the RS and LS contacts exhibit distinct characteristics. The RS spectrum features a prominent peak at $B= -0.25$~T, accompanied by a small shoulder peak around $-0.18$~T, consistent with the presence of two Fermi surfaces. In contrast, the LS spectrum displays a singular, broad peak at $-0.21$~T. 

To understand the experimental data, we simulate the TMF spectra of holes as shown in Figure 3e. The simulation provides insights into the features observed in the experimental TMF spectra. Specifically, the small shoulder at $-0.18$~T in the RS spectrum is attributed to the focusing of $\beta_{\mathrm{in}}$ holes onto the collector, as evidenced by the trajectory simulation in Figure 3f. Meanwhile, the more pronounced peak is linked to the focusing of $\beta_{\mathrm{out}}$ holes. In the case of the LS spectrum, the simulation reveals that the focusing fields for $\beta_{\mathrm{in}}$ and $\beta_{\mathrm{out}}$ are closely matched (Figure 3e, bottom), causing the two peaks to fully merge into a single peak, consistent with our experimental data. The primary contributing factor to the broadening of the focusing peaks in our device stems from the finite width of the injector and collector, approximately $500$~nm, leading to poorly resolved focusing peaks. This observation underscores the importance of considering spatial constraints in the design and interpretation of TMF spectra in nanoscale devices.

\begin{figure*}
  \includegraphics{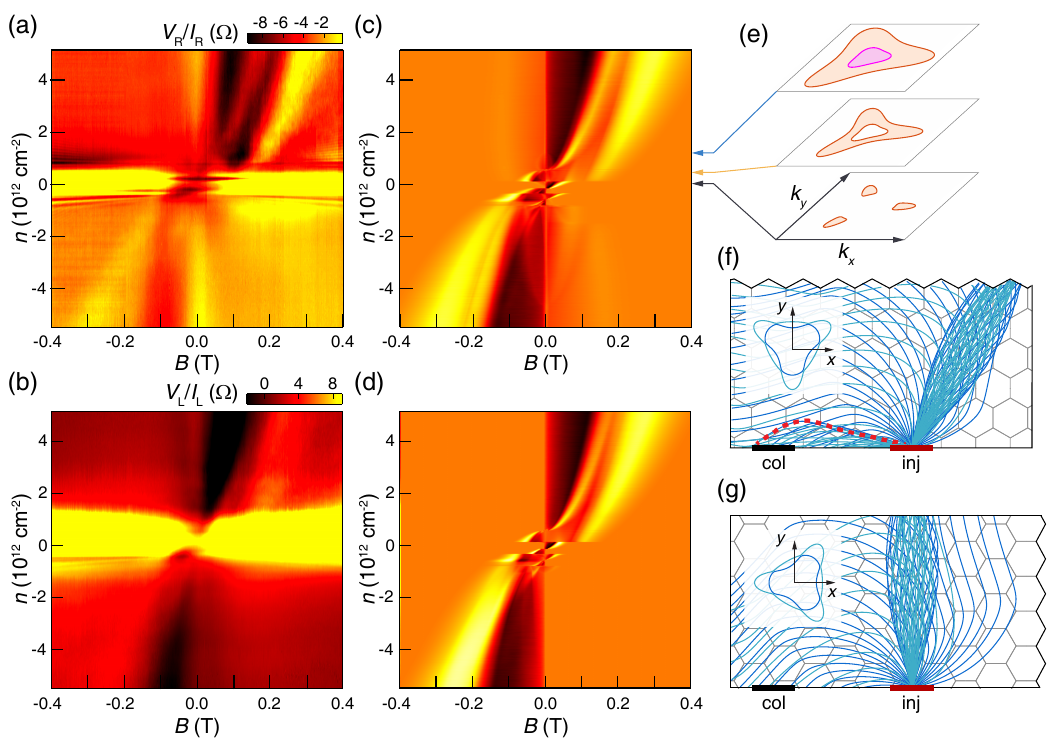}
  \caption{(a, b) Experimental TMF spectra at $2.4$~K for RS and LS contacts, respectively. (c, d) Numerical simulation of the TMF data for RS and LS contacts. (e) Fermi surfaces at densities $1.2\times 10^{12}$ (blue), $4.5 \times 10^{11}$ (yellow), and $1.0 \times 10^{11}$~cm$^{-2}$ (black). (f, g) Simulated trajectories of holes at $n = -2\times10^{12}$ $\mathrm{cm^{-2}}$ and $B=-0.05$ T for RS and LS contacts. The red dash line in (f) shows a caustic. The inset shows real-space trajectories of holes in each condition.}
\end{figure*}

We further investigate the density dependence of TMF spectra. Figure 4a and 4b show the RS and LS TMF spectra as a function of $B$ and $n$. The focusing peaks are represented as bright curved ridges along diagonal in the plot. At high carrier density ($|n| > 1\times10^{12}$ $\mathrm{cm^{-2}}$), the TMF spectra exhibit behavior consistent with those presented in Figure 3. The RS spectra manifest two distinct peaks, whereas the LS spectra display poor resolution due to the smaller difference in $B_{\mathrm{f}}$ between the two Fermi surfaces. As $n$ decreases, $B_{\mathrm{f}}$ for each pocket also decreases. This behavior arises from the shrinking Fermi surfaces at lower carrier densities, resulting in a smaller cyclotron radius of carriers. The $n$-dependent TMF spectra for both sides at high density align well with the simulation, as demonstrated in Figure 4c and 4d. This agreement further underscores the robustness and reliability of our experimental observations and simulations across varying carrier densities. 

As the electron density reduces below approximately $1\times10^{12}~\mathrm{cm^{-2}}$, the inner peak associated with the $\alpha_{\mathrm{in}}$ band disappears (Figure 4a), in agreement with our simulation, depicted in Figure 4c. The vanishing of this inner peak results from the Fermi level dropping below the bottom edge of the $\alpha_{\mathrm{in}}$ band. At these lower densities, only electrons from the $\alpha_{\mathrm{out}}$ band contribute, resulting in a Fermi surface that includes both electron and hole orbits, as shown in Figure 4e (highlighted in yellow). Our simulation reveals the coexistence of electron and hole orbits within a certain density range, leading to the appearance of peaks at both positive and negative $B$ values. As $n$ approaches zero, the simulation shows a pronounced shift in the focusing field of the outer peak, indicative of a Lifshitz transition. This transition marks a topological transformation of the Fermi surface from a singular to three distinct smaller electron pockets (Figure 4e, black). However, in our experimental data, this Lifshitz transition is not directly observable as it occurs at densities akin to disorder-induced density fluctuations (the horizontal ridge in Figure 4a).

The simulation further unveils an additional focusing peak, as a small shoulder in the RS data near $B = -0.05$~T and $n = -2\times 10^{12}$~cm$^{-2}$ (Figure 4a). This supplementary peak, elucidated by the simulation, emerges due to the specific injection direction of holes in a device oriented along the zigzag direction. Figure 4f illustrates this phenomenon, where a new caustic generates a focusing path onto the collector. Notably, this extra peak is absent in the LS data corresponding to the armchair orientation. The simulation of carrier trajectories in Figure 4g aligns with this observation, as the caustic does not direct carriers toward the collector in this particular configuration. These findings underscore the orientation-dependent nuances in the TMF behavior of anisotropic Fermi surfaces.

In summary, we employed TMF to investigate the anisotropic Fermi surface in 4LG. Our findings reveal that TMF spectra distinctly vary with crystal orientations, a critical observation that underscores the sensitivity of TMF in detecting subtle electronic structure changes linked to TW effects. This variation in TMF spectra not only deepens the understanding of anisotropic properties in 4LG but also illustrates TMF's potential in elucidating the complex relationship between crystal orientation and anisotropic electronic properties in multilayer graphene. These insights have broader implications for materials with anisotropic Fermi surfaces, expanding the scope of TMF applications in material science research.

\begin{acknowledgments}
This research has been primarily supported by the NSRF via the Program Management Unit for Human Resources \& Institutional Development, Research and Innovation (Grant no. B05F640152), National Research Council of Thailand (NRCT) and Chulalongkorn University (Grant no. N42A650266), and the Thailand Toray Science Foundation (TTSF). K.W. and T.Taniguchi acknowledge support from the JSPS KAKENHI (Grant Numbers 20H00354 and 23H02052) and World Premier International Research Center Initiative (WPI), MEXT, Japan.
\end{acknowledgments}

%\bibliography{ref}


\newpage
\onecolumngrid
%\appendix
\renewcommand{\theequation}{S\arabic{equation}}
\renewcommand{\thefigure}{S\arabic{figure}}
\renewcommand{\thetable}{S\arabic{table}}
\renewcommand{\theenumi}{S\arabic{enumi}}
\renewcommand{\thesection}{S\Roman{section}}
\setcounter{figure}{0} 
\begin{center}
	\LARGE{\bf Supporting Information}
\end{center}

{\large\bf Device fabrication}

\emph{2D materials preparation.} Tetralayer graphene (4LG) and hexagonal boron nitride (hBN) flakes were exfoliated from their bulk counterparts using the scotch tape technique \cite{huang2015reliable} and subsequently deposited onto Si/SiO\textsubscript{2} chips. The optical contrast between the flakes and SiO\textsubscript{2} substrates was used to identify the thickness of the 2D materials \cite{wang2012thickness}. Flakes of 4LG with perpendicular cleaved edges and hBN flakes with a thickness of approximately 10-30 nm were carefully selected from the chips for device fabrication.  The hBN and 4LG flakes used for the device in the main text are depicted in Figure S1(a-c).\\

\emph{Heterostructure assembly.} To create the hBN/4LG/hBN heterostructure (BNGBN), a stack comprising poly(bisphenol A carbonate) film on a block of polydimethylsiloxane (PDMS) was utilized as a stamp. This stamp was employed in picking up the selected flakes and precisely stacking them together. Subsequently, the BNGBN heterostructure was released onto a new Si/SiO\textsubscript{2} chip at 180 $^{\circ}$C to eliminate blisters that may have been trapped between the interfaces of the heterostructure \cite{purdie2018cleaning}. Figure S1(d) illustrates the optical image of the heterostructure following the assembly process.\\

\emph{Electron-beam lithography.}  The blister-free region within the BNGBN (marked by the dashed line in Figure S1(d)) was used as the basis for designing the transport channel for the device discussed in the main text. A polymethyl methacrylate (PMMA) film was used as the electron beam resist. Subsequently, EBL and O\textsubscript{2}/CHF\textsubscript{3} plasma etching techniques were employed to define an L-shaped device, with two edges of the device aligned parallel to the cleaved edge of the 4LG. In Figure S1(e), the device structure is shown, indicating the deposition of the top gate and contacts using thermal evaporation of Cr/Au at a thickness of 5/50 nm.\\

\emph{Control of charge carrier density.} The L-shaped device depicted in Figure S1(e) comprises a p-doped silicon back gate (BG) and a Cr/Au top gate (TG). By applying voltages $V_{\mathrm{BG}}$ and $V_{\mathrm{TG}}$ to the BG and TG, we can introduce a carrier density $n = \frac{1}{e}(C_{\mathrm{BG}}V_\mathrm{BG} + C_{\mathrm{TG}}V_{\mathrm{TG}})$ and an electric displacement field $D = \frac{1}{2\epsilon_{0}}(C_{\mathrm{BG}}V_\mathrm{BG} - C_{\mathrm{TG}}V_{\mathrm{TG}})$ into the 4LG channel. Here, $e$ is the elementary charge, $\epsilon_{0}$ is the vacuum permittivity, and $C_{\mathrm{BG}}$ and $C_{\mathrm{TG}}$ are the capacitance per unit area of the BG and TG, respectively (determined from Shubnikov-de Haas oscillations at finite magnetic fields). Based on these relationships, we can introduce charge carriers into the 4LG without an external displacement field by simultaneously adjusting both $V_{\mathrm{BG}}$ and $V_{\mathrm{TG}}$ in a manner that satisfies the relation $C_{\mathrm{TG}}V_{\mathrm{TG}} = C_{\mathrm{BG}}V_{\mathrm{BG}}$.\\

\begin{figure}
  \centering
  \includegraphics[width=\textwidth]{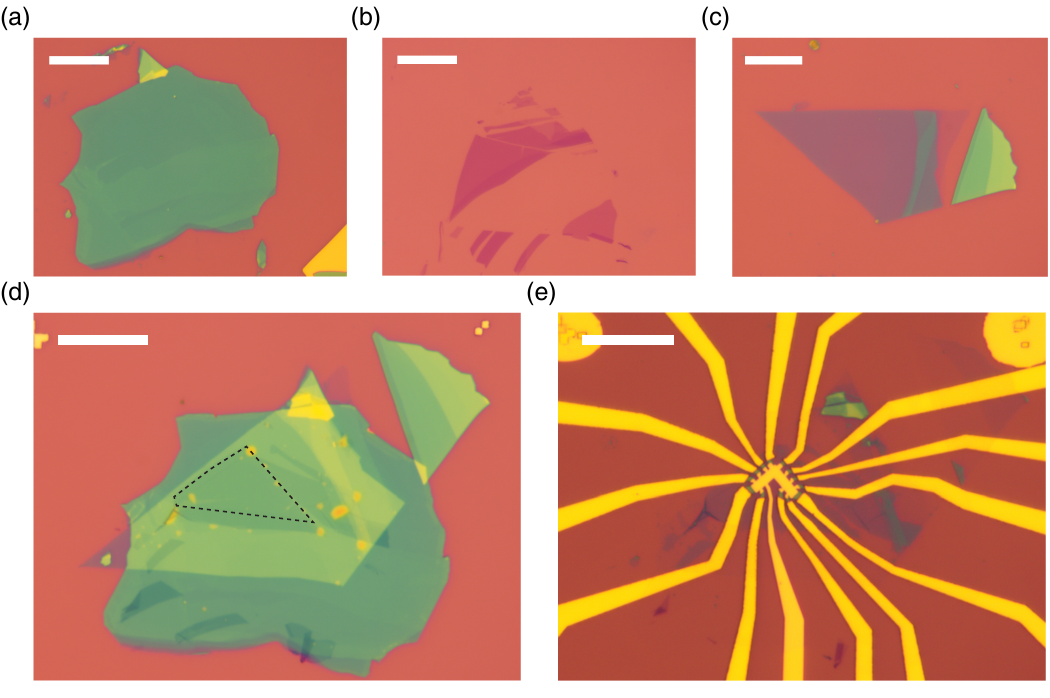}
  \caption{(a-c) Individual top hBN (a), 4LG (b), and bottom hBN (c) flakes selected in constructing the device discussed in the main text. (d) The hBN/4LG/hBN heterostructure formed from these flakes (a-c). The black dashed line indicates the edge of 4LG. (e) The device following the lithography processes. Scale bars are 20 $\mu$m.}
\end{figure}

{\large \bf Local transport measurement}

\begin{figure}
  \centering
  \includegraphics[width=\textwidth]{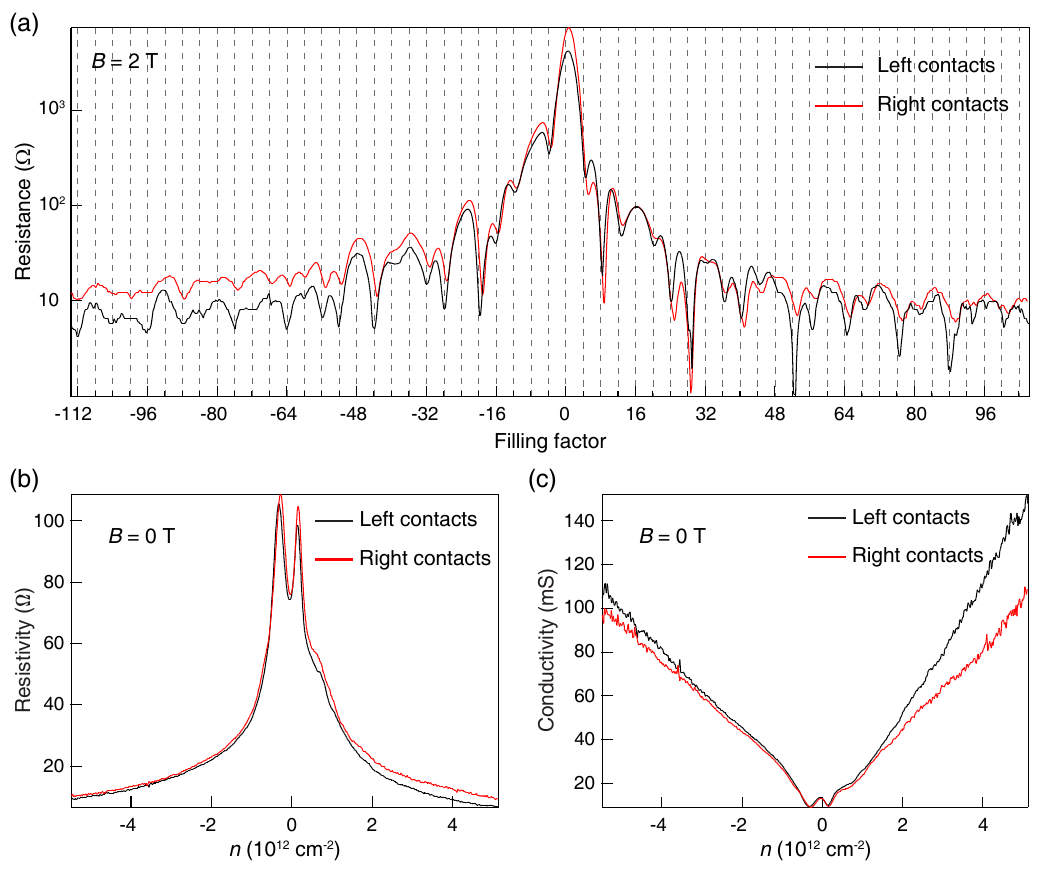}
  \caption{(a) Four-probe longitudinal resistance as a function of filling factor at $B$ = $2$~T and $2.4$~K. The black and red lines are measured using contacts from the left and right sides of the device, respectively. (b) Resistivity and (c) conductivity as functions of $n$ at $B$ = 0 T. The black and red lines represent the data from left and right contacts, respectively}
\end{figure}

The stacking configuration of our 4LG device was determined by analyzing the Landau-level (LL) crossings in the quantum Hall data. As illustrated in Figure 2d of the main text, the longitudinal resistance for the left-side contacts reveals the positions of these LL crossings. Notably, we observed that these LL crossing positions align precisely with those reported in Ref. \citenum{shi2018tunable}, confirming the ABAB stacking of our device. To ensure uniform stacking across the entire device, we compared the resistance oscillations due to the Shubnikov-de Hass effect between both sides. Figure S2 presents the resulting data depicting the resistance oscillations for the left-side (black line) and right-side (red line) contacts under a perpendicular magnetic field $B$ = 2 T as a function of filling factor $\nu$, with a fixed current bias of 300 nA applied to both sides. Notably, dips in the resistance correspond to integer values of $\nu$, with the majority of these resistance dips exhibiting a separation of $\Delta \nu$ = 4, attributed to spin and valley degeneracy. However, certain resistance dips, such as those at $\nu$ = 12 and 20, are separated by $\Delta \nu$ = 8 due to Landau level crossings. Given that the LL crossing positions are consistent for both sides, we conclude that our 4LG device maintains a uniform Bernal stacking configuration. Therefore, the observed variation in the TMF spectra in the main text is not attributable to differing stacking orders of 4LG between the left- and right-side contacts.

To evaluate the quality and density uniformity of the 4LG across both the left and right sides of the device, we conducted resistivity measurements as functions of carrier density ($n$) at $B = 0$ T. The resistivity profiles for both sides, depicted in Figure S2(b), exhibit characteristic double peaks typical of ABAB-stacked 4LG \cite{shi2018tunable}. Conductivity, illustrated in Figure S2(c), was derived from the reciprocal of resistivity for the left and right contacts. Notably, on the left side of the device, field-effect mobilities, determined from the linear regime of the conductivity plots, were identified to be 180,000 and 110,000 $\mathrm{cm^2V^{-1}s^{-1}}$ for electrons and holes, respectively. Conversely, the right side exhibited slightly lower mobilities of 110,000 and 100,000 $\mathrm{cm^2V^{-1}s^{-1}}$ for electrons and holes, respectively. These notably high mobility values on both sides suggest minimal potential fluctuation.

\newpage 

{\large \bf Band structure calculation}

The low-energy band structure of tetralayer graphene (4LG) near the $K$ and $K'$ points in the graphene Brillouin zone was computed by diagonalizing the effective Hamiltonian, which was parameterized using the Slonczewski-Weiss-McClure parameters of graphite. The effective Hamiltonian follows the details in Ref. \citenum{koshino2011landau}. Our chosen basis consisted of $|A_1\rangle$, $|B_1\rangle$, $|A_2\rangle$, $|B_2\rangle$, $|A_3\rangle$, $|B_3\rangle$, $|A_4\rangle$, and $|B_4\rangle$, where $|A_i\rangle$ and $|B_i\rangle$ represent Bloch states for electrons in layer $i$ on sublattices $A$ and $B$, respectively. In this basis, the Hamiltonian for the 4LG can be written as:  

\[ H_{\mathrm{4LG}} = 
\begin{pmatrix}
    0                   & v\pi^{\dagger}    & -v_{4}\pi^{\dagger}   & v_{3}\pi              & \gamma_{2}/2          & 0                 & 0                 & 0 \\
    v\pi                & \Delta'           & \gamma_{1}            & -v_{4}\pi^{\dagger}   & 0                     & \gamma_{5}/2      & 0                 & 0 \\
    -v_{4}\pi           & \gamma_{1}        & \Delta'               & v\pi^{\dagger}        & -V_{4}\pi             & \Gamma_{1}        & \gamma_{5}/2      & 0 \\
    v_{3}\pi^{\dagger}  & -v_{4}\pi         & v\pi                  & 0                     & V_{3}\pi^{\dagger}    & -V_{4}\pi         & 0                 & \gamma_{2}/2 \\
    \gamma_{2}/2        & 0                 & -V_{4}\pi^{\dagger}   & V_{3}\pi              & 0                     & v\pi^{\dagger}    & -v_{4}\pi^{\dagger} & v_{3}\pi \\
    0                   & \gamma_{5}/2      & \Gamma_{1}            & -V_{4}\pi^{\dagger}   & v\pi                  & \Delta'           & \gamma_{1}        & -v_{4}\pi^{\dagger} \\
    0                   & 0                 & \gamma_{5}/2          & 0                     & -v_{4}\pi             & \gamma_{1}        & \Delta'           & v\pi^{\dagger} \\
    0                   & 0                 & 0                     & \gamma_{2}/2          & v_{3}\pi^{\dagger}    & -v_{4}\pi         & v\pi              & 0
\end{pmatrix},
\]

\noindent where $\Delta'$ is the energy difference between dimer and non-dimer sites, $\gamma_{i}$ are the hopping parameters as indicated by the diagram in Figure 2f of the main text, and the velocities $v_{i} = \sqrt{3}a\gamma_{i}/2\hbar$ (the lattice constant $a$ = 0.246 nm). Using the convention that the x-axis is parallel to 4LG's zigzag edge, we can express $\pi = \xi p_{x} + i p_{y}$, where $\mathbf{p} = (p_{x}, p_{y})$ is the momentum operator and $\xi = \pm 1$ is the valley index.

In the Hamiltonian above, we have introduced an additional set of hopping parameters $\Gamma_{1}$, $\Gamma_{3}$, and $\Gamma_{4}$, along with their corresponding velocities $V_{i} = \sqrt{3}a\Gamma_{i}/(2\hbar)$, to account for difference in interlayer interactions between the bulk and surfaces of 4LG \cite{wu2015detection}. Diagonalizing this Hamiltonian at each momentum $\mathbf{p} = \hbar\mathbf{k}$ using the same parameters as in Ref.\citenum{shi2018tunable} yields the low energy band structure as depicted in Figure 1a of the main text. \\
\\

{\large \bf TMF spectra simulation}
\begin{figure}
  \centering
  \includegraphics[width=\textwidth]{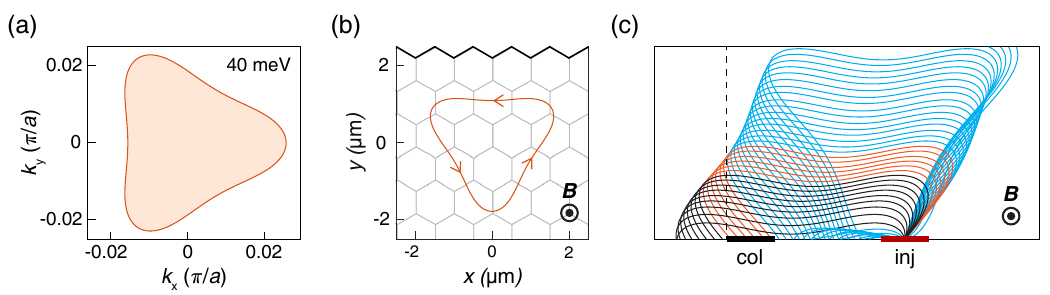}
  \caption{(a) Fermi surface of 4LG from band $\alpha_{\mathrm{out}}$ at $E_{\mathrm{F}} = 40$ meV. (b) Trajectory of $\alpha_{\mathrm{out}}$ electrons in real space with $E_{\mathrm{F}} = 40$ meV under an out-of-plane magnetic field $B$ = 0.12 T. (c) Simulated trajectories of $\alpha_{\mathrm{out}}$ electrons, with the same conditions as in (b), injected at the red contact from various initial angles. Orange and black paths represent trajectories reaching the collector and reference contacts, respectively. The zigzag edges of the 4LG in (b) and (c) are aligned parallel to the x-axis.}
\end{figure}

In our simulation of the TMF spectra, presented in Figure 3 of the main text, we initiated the process by deriving the Fermi surface from the band structure, corresponding to a specific Fermi energy, denoted as $E_{\mathrm{F}}$. This is depicted in Figure S3(a), where we illustrate the Fermi surface associated with electrons in the $\alpha_{\mathrm{out}}$ band at an $E_{\mathrm{F}}$ value of 40 meV. When subjected to a magnetic field $B$, we determine the semiclassical trajectory by implementing a 90$^{\circ}$ rotation of the Fermi surface. This rotation is clockwise for electrons and counter-clockwise for holes. Additionally, the Fermi surface is scaled by a factor of $\hbar/eB$, as demonstrated in Figure S3(b). The direction of motion within the orbit is influenced by the effective charge of the carriers and the direction of magnetic field.

In our study, the non-local resistance for each $B$ is estimated to be directly proportional to the ratio $(N_{\mathrm{c}} - N_{\mathrm{ref}})/N_{\mathrm{i}}$. The simulation begins with the modeling of electron trajectories, injected at various initial angles through the injector, as depicted in Figure S3(c). These trajectories are monitored to quantify the fraction that successfully arrives at the collector. To accurately represent the density of states at each angle, each trajectory is assigned a weight inversely proportional to $|\nabla_{\mathbf{k}} \epsilon(\mathbf{k})|$. The total number of injected electrons, denoted as $N_{\mathrm{i}}$, is calculated by summing the weights of all trajectories corresponding to injected electrons. Conversely, the number of electrons that reach the collector, $N_{\mathrm{c}}$, is determined by summing the weights of those trajectories that successfully terminate at the collector, as shown by the orange paths in Figure S3(c).

To estimate the number of electrons reaching the reference contact, denoted as $N_{\mathrm{ref}}$, we adopt the premise that only those electron trajectories which extend to the device's edge beyond the collector contact are relevant contributors to $N_{\mathrm{ref}}$. These trajectories are represented by the black paths in Figure S3(c). Given the absence of second-order focusing peaks in our TMF spectra, it is inferred that diffusive scattering occurs at the device's edge. Consequently, trajectories that scatter off the device's edge in the region between the injector and collector (illustrated as blue paths in Figure S3(c)) are presumed to reach both the collector and the reference contact equally. Therefore, these do not influence the TMF spectra. The value of $N_{\mathrm{ref}}$ is derived by aggregating the weights of the black trajectories depicted in Figure S3(c).

%\bibliography{ref}

\end{document}